# Parameterized Type Definitions in Mathematica: Methods and Advantages

Alina Andreica

Faculty of European Studies, "Babeş-Bolyai" University of Cluj-Napoca, Romania
E-mail: alina.andreica@euro.ubbcluj.ro

**Abstract**. The theme of symbolic computation in algebraic categories has become of utmost importance in the last decade since it enables the automatic modeling of modern algebra theories. On this theoretical background, the present paper reveals the utility of the parameterized categorical approach by deriving a multivariate polynomial category (over various coefficient domains), which is used by our Mathematica implementation of Buchberger's algorithms for determining the Gröbner basis. These implementations are designed according to domain and category parameterization principles underlining their advantages: operation protection, inheritance, generality, easy extendibility. In particular, such an extension of Mathematica, a widely used symbolic computation system, with a new type system has a certain practical importance. The approach we propose for Mathematica is inspired from D. Gruntz and M. Monagan's work in Gauss, for Maple.

## 1. Theoretical Background: the Type Theory in Symbolic Computation

The type theory in symbolic computation [San93], [San95], [San96], [Dav90], [Dav91] lies between the mathematical category theory and the type theory from programming languages, extended with the inheritance concept, according to object oriented (OO) principles. The final goal of this theory is to create techniques for an efficient and systematic automatic implementation of symbolic computations within various algebraic domains, including modern algebra structures. Thus, there can be defined a hierarchy of algebraic structures, within which common mathematical domains are appropriately derived.

The most important concepts in this type theory are domains and categories, which are borrowed from algebra and correspond, in OO programming, to classes, respectively metaclasses. Therefore, the algebraic structures and the relationships between them will be modeled by means of classes, subclasses, instances and the inheritance concept. The necessity of adopting some other approach than the theoretical algebraic one relies first of all on the constructive and algorithmic character of type theory, where new types, characterized by the same properties and methods, are introduced in a functional manner, using existing types [And98]. The types defined in this manner create general contexts of expressing algorithms, which can therefore be applied for various types of data (similar to generic programming). These general structures are named categories and the more specific types generated by them are named domains.

Unifying the algebraic (see [Pur82] for specific concepts, i. e. category) and object oriented (see [FrePar94] for specific concepts, i. e. data type, instance, inheritance etc.) points of view, we propose the following definitions for type theory concepts:

**Definition 1**. In symbolic computation, we define a ***category*** as a data type which describes an algebraic category.

**Definition 2**. In symbolic computation, we define a ***domain*** as a data type that instantiates (in OO sense) a category (see definition 1).

**Definition 3**. A ***package*** is a program unit which describes the behavior (specific processing algorithms) of a data collection.

**Definition 4**. A ***parameterized*** data type is a data type that depends on another data type, given as parameter.

Based on the type theory concepts, algebraic structures can be defined pursuing systematic, functional and algorithmic principles, which would facilitate their implementation into programming languages [And99]. For each type of algebraic structure, characteristic operators (composition operator, equality operator, symmetric operator, or other operators that describe constructively certain mathematical properties) are defined in a functional manner. The operators are associated additional axioms, such as associativity, commutativity or distributivity). Each type of algebraic structure inherits, from its ascendant, the corresponding operators, and supplementary introduces only its specific operators and axioms.

We implemented a part of the algebraic structures' hierarchy, containing semigroups, monoids, abelian monoids, groups, abelian groups, rings, abelian rings, fields and abelian fields, as an extension of Mathematica with a new type system [And97] – *HierMath* package. This package introduces, into a SCS that is widely used for computations in common mathematical domains, new facilities, regarding the definition and computational use of abstract algebraic domains.

The goal of the present paper is to prove the utility of type theory principles, in particular of parameterized definitions [GruMon93] of categories and domains, by deriving a multivariate polynomial category, whose parameterized definition is used in the construction of some specific polynomial domains, over various coefficient rings. These polynomial domains are used by the algorithms which implement Buchberger's method for computing the Gröbner basis and the reduced Gröbner basis for a given set of polynomials [Buc85], [Gru93]. The efficiency of the implementation we propose consists in its generality and in the fact that is easily extendable by defining new domains, which can be specified as parameters in the same category definition. As far as we know, this type of implementation is a novelty for Mathematica; similar approaches were proposed in Gauss, for Maple system [Gru93], [GruMon93]. It fact, these works inspired the present approach.

## 2. Parameterized Definitions of Polynomial Domains

Implementing symbolic computations from the category theory point of view brings in symbolic computations advantages that are similar to the ones introduced by object oriented (OO) principles in programming. Thus, there can be defined operations (methods) specific to certain working contexts, operations that will be protected to inconsistent accesses (similar to classes in OO programming). Moreover, the implementation of the category / domain hierarchy, is simplified by using inheritance principles. The hierarchic and parameterized methods of definition have the advantage of easily extending the system of algebraic structures, on the same general definition principles.

We intend to exemplify these ideas by defining a multivariate polynomial category, which will be parameterized with various coefficient domains ($\wedge$, $\wedge$ mod n, squared matrices). On the same parameterized principles [GruMon93], when a new coefficient domain will be defined, the same polynomial category will be used, by only modifying its parameter for the new coefficient domain.

### 2.1. Defining the Exponent Vector Domain

With the view to implementing the multivariate polynomial category, we define an auxiliary domain that will manipulate monomials, naming it the exponent vector domain [Gru93]. For a given base of identifiers (for example, {x,y,z}), an exponent vector will be retained by the list of exponents corresponding to each variable. In operating upon exponent vectors, we use two types of representations: lists, respectively products of primes with the exponents in question [Gru93]. Thus, using the latter representation, the exponent vector {$e_1$, $e_2$, …, $e_n$} will be retained and processed with the number $p_1^{e_1} p_2^{e_2} ... p_n^{e_n}$, where $p_1$, $p_2$, …, $p_n$ are prime numbers, for simplifying computations – the first prime numbers. In this representation, an exponent vector sum reduces to the corresponding prime numbers product, whereas the greatest common divisor (Gcd) and the least common multiple (Lcm) can be computed by similar operations upon prime products.

The exponent vector domain is an abelian monoid [And97] that introduces the following operations:

♦ computing neutral, minimum and maximum elements (“`0`”, “`Max`”, “`Min`”);
♦ conversions between the list and number internal forms (“`ListaInVectorNr`”, “`VectorNrInLista`”);
♦ sum (“`+`”), greatest common divisor (“`Gcd`”) and least common multiple (“`Lcm`”) for two exponent vectors (we mention that “`GCD`” and “`LCM`” are alternative definitions, using different representations);
♦ positiveness test for an exponent vector (“`Pozitiv`”);

- divisibility test for two exponent vectors ("|");
- relational operators ("<", ">", "=", "<=", ">=", "<>") between two exponent vectors – we shall consider the *lexicographical ordering* (corresponding relations are implemented by `string[…]` functions);
- conversions between external and internal forms ("Inp", "Out").

We give below the most relevant part of Mathematica code which defines the exponent vector domain (italics are used for functions with omitted bodies). All operations within an exponent vector domain `V`, created by the function `VectExp[V,lv]`, where `lv` is the variable list representing the base, will be prefixed with the domain name and will have as the first parameter the operation code. For example, `V["+",v1,v2]` returns the sum of two exponent vectors (in internal form), `V["Gcd",v1,v2]` computes the greatest common divisor, `V["Lcm ",v1,v2]` computes the tleast common multiple, etc.

```
BeginPackage["VectoriExp`"]

VectExp::usage="VectExp[V_,lv_List] defineste domeniul V de vectori de
  exponenti cu baza lv"
   (*defines the exponent vector domain V, with the base lv*)
Intreg::usage="Domeniu intreg" (*integer domain*)
Primi::usage="lista de numere prime" (*list of first primes*)
string::usage="operatii pe domeniul String" (*string operations*)

Begin["VectoriExp`Private`"]
Needs["HierMath`"]

string["Rel", s1_String, s2_String]:=Module[{i,rel,l1,l2,nr,n1,n2},
 (*tests the relation <, =, > between s1 and s2, returning -1,0,1*) …]
 string["<",s1_String, s2_String]:=string["Rel", s1, s2]==-1;
 string[">",s1_String, s2_String]:=string["Rel", s1, s2]==1;
 string["=",s1_String, s2_String]:=string["Rel", s1, s2]==0;
 string["<>",s1_String, s2_String]:=string["Rel", s1, s2]!=0;

Primi={2,3,5,7,11,13,17,19,23,29,31}
Intreg[Max]=Min[]
Intreg[Min]=Max[]

VectExp[V_,lv_List]:=Module[{n,en,k,max,min},
 (*creates the exponent vector domain V, with the base lv*)
n=Length[lv];
 For[k=1;en={};max={};min={}, k<=n, k++,
     en=Append[en,0]; max=Append[max,Intreg[Max]];
     min=Append[min,Intreg[Min]]];
 MonoidCom[V,"+",en];                (*creates an abelian monoid [And97]*)

 V["0"]=en;

 V["ListaInVectorNr",l_List]:=Module[{i,nr,p},    nr=Length[l]; p=1;
  Do[p*=Primi[[i]]^l[[i]], {i,nr}];
  Return[p] ];
```

```
V["VectorNrInLista",nr_Integer]:=Module[{x,i=1,aux,l={}},  x=nr;
 While[x>1,
   d=Primi[[i]]; aux=0;
   While[Mod[x,d]==0, aux+=1; x=Quotient[x,d]];
   If[aux!=0, l=Append[l,aux],];
   i=i+1];
 Return[l]  ];

V["+",l1_List,l2_List]:=l1+l2;
V["-",l1_List,l2_List]:=l1-l2;

V["Pozitiv",l_List]:=
  Module[(*tests whether all elements in the list are >0*) …]

V["Gcd", l1_List, l2_List]:= V["VectorNrInLista",
   GCD[V["ListaInVectorNr",l1], V["ListaInVectorNr",l2]]];
V["GCD",l1_List, l2_List]:=MapThread[Min,List[l1,l2]];

V["Lcm", l1_List, l2_List]:=V["VectorNrInLista",
  LCM[V["ListaInVectorNr",l1], V["ListaInVectorNr",l2]]];
V["LCM",l1_List,l2_List]:=MapThread[Max,List[l1,l2]];

V["Rel", l1_List, l2_List]:=Module[{i,rel}, i=1; rel=0;
 (*tests the relation <, =, > between l1 and l2, returning -1,0,1*) …]
V["<",l1_List,l2_List]:=V["Rel", l1, l2]==-1;
V[">",l1_List,l2_List]:=V["Rel", l1, l2]==1;
V["=",l1_List,l2_List]:=V["Rel", l1, l2]==0;
V["<>",l1_List,l2_List]:=V["Rel", l1, l2]!=0;
V["<=",l1_List,l2_List]:=V["<",l1,l2] || V["=",l1,l2];
V[">=",l1_List,l2_List]:=V[">",l1,l2] || V["=",l1,l2];

V["|",l1_List,l2_List]:=Module[{n1,n2,i,bool},
  n1=Length[l1]; n2=Length[l2];     bool=n1<=n2;
  For[i=1,i<=Min[n1,n2],i++,  If[l1[[i]]>l2[[i]], bool=False,]];
  Return[bool];  ];

V["Out",l_List]:=Module[{lr,i},  lr={};
 For[i=1,i<=n,i++,  lr=Append[lr,Apply[Power, List[lv[[i]],l[[i]]]]]];
 Return[Apply[Times,lr]]];

V["Inp",e__]:=Module[…
  (*transforms an input with the syntax x[^e1]*y[^e2]... into the
   internal list form; the code is rather complex and based on
   Mathematica internal forms*)  …   ]
V["Max"]:=max;
V["Min"]:=min;]

End[]
EndPackage[]
```

## 2.2. Defining A Multivariate Polynomial Category over Various Coefficient Domains

In order to operate upon multivariate polynomials, we created two implementations and studied their efficiency, compared to Mathematica built-in facilities.

The *first representation* (**polinom.m** package) views a polynomial as a list of two elements: the exponent vector list, <u>lexicographically ordered</u>, and the corresponding coefficient list. For example, the polynomial $2*x\string^2*z-5*y$ (with the base $\{x,y,z\}$) will be rezprezented as $\{\{\{0,1,0\},\{2,0,1\}\}, \{-5,2\}\}$.

The *second representation* (**polin.m** package) views a polynomial as a table T which retains in T[0] the exponent vctor list, <u>lexicographically ordered</u>, and the coefficients of each exponent vector V (represented as a list) – in T[V]. For example, the polynomial $2*x\string^2*z-5*y$ (with the base $\{x,y,z\}$) will be represented by a table P with: P[0]=$\{\{0,1,0\},\{2,0,1\}\}$, P[$\{0,1,0\}$]=-5, P[$\{2,0,1\}$]=2.

In order to simplify the manipulation of the exponent vector list, we provide it with a sentinel, which contains maximum exponent values (over the appropriate domain).

The polynomial category is an abelian ring parameterized with the coefficient and exponent vector domains (the latter – over a certain variable base); this category will generate, for different parameters, particular polynomial domains. For each of the two implementations, we defined the following operations:

♦ selecting the current coefficient and exponent vector domains (`"DomCoef"`, `"DomVectExp"`);

♦ initializing a polynomial (`"Init"`) and testing null polynomials (`"Nul"`);

♦ for a given polynomial, computing the dimension (`"Nr"`), the "leading" monomial – according to the lexicographical ordering (`"MonomGrMax"`), the "leading" coefficient (`"CoefGrMax"`), the monomial of a certain index (`"Monom"`), the coefficient of a certain index (`"Coef"`) and the position of a certain monomial (`"Indice"`);

♦ adding a monomial with a certain coefficient to a given polynomial (`"AdaugMonom"`);

♦ converting a monomial into a polynomial (`"MonomInPol"`) and copying a polynomial (`"Copy"`);

♦ for two given polynomials, computing the sum ("+"), product ("*") and subtraction ("−");

♦ multiplying a polynomial by a number (`"&"`);

♦ dividing a polynomial by a given monomial ("/") and performing the divisibility test for a polynomial and a monomial ("|");

♦ conversions between the internal and external forms of a polynomial (`"Inp"`, `"Out"`).
<u>Observation</u>. In implementing these functions we supposed that coefficients are elementary, but the algorithm is easily expandable for list type coefficients (for example, matrices).

We give below the main part of the Mathematica package which defines the polynomial category, using the first representation from the ones described above (**polinom.m** package; italics are used for functions with omitted bodies). The second implementation is similar. Within Mathematica code, one can notice the functional and parametric specification of operations within various domains. For example, within the polynomial domain Pol, defined by `Polinom[Pol, DCoef, DVect, l]`, where

`DCoef` is the coefficient domain and `DVect` is the exponent vector domain with the base `l`, the construction `DVect["|",v,Pol["Monom",i,P]]` performs a divisibility test – within `DVect` exponent vector domain – between two exponent vectors, the second one corresponding to a monomial selected from a polynomial `P` (by "Monom" operation, within the polynomial domain `Pol`). `DVect["+",o1,o2]` is the sum of `o1` and `o2` in `DVect` domain, whereas `DCoef["+",o1,o2]` is a sum in `DCoef` domain.

```
BeginPackage["Polinom`"]

Polinom::usage="Polinom[Pol,DCoef,DVect,l] defineste domeniul Pol de
  polinoame de mai multe variabile, peste domeniile DCoef pentru
  coeficienti si DVect pentru vectorii de exponenti (monoame), cu baza
l"

Begin["Polinom`Private`"]
Needs["VectoriExp`","DomCoef`"]

Polinom[Pol_,DCoef_,DVect_,baza_List]:=Module[{n,lv},
  (*defines a multivariate polynomial domain, with coefficients in
  DCoef, over the exponent vector domain Dvect, with the base baza*)

 InelCom[DCoef,"+","*"];            (*creates an abelian ring [And97]*)
 VectExp[DVect,baza];              (*see 2.1*)

 Pol["DomCoef"]:=DCoef;
 Pol["DomVectExp"]:=VectExp;
 Pol["Init"]:=List[List[DVect["Max"]],List[]];  (*returns the null
   polynomial, with a sentinel in the exponent vector list*)
 Pol["Nul",P_]:=SameQ[P[[1]],List[DVect["Max"]]];
  (* returns True if P is null *)
 Pol["Nr",P_]:=Length[P[[1]]]-1; (*dimension*)
 Pol["Monom", i_,P_]:=P[[1]][[i]];
    (*the ith monomial from the polynomial P*)
 Pol["Coef",i_,P_]:=If[i<=Pol["Nr",P],P[[2]][[i]],0];
 Pol["Indice",l_,P_]:=Position[P[[1]],l,1][[1,1]];
  (*the index of the monomial l in the polynomial P*)
 Pol["MonomGrMax",P_]:=P[[1]][[Pol["Nr",P]]];
 Pol["CoefGrMax",P_]:=P[[2]][[Pol["Nr",P]]];

 Pol["AdaugMonom",T_,v_,c_]:=Module[{nr,i,k,l,lc,coef},
  (*adds to the polynomial T the monomial formed by the exponent
    vector v and the coefficient c taking into account the
    lexicographical ordering; if v exponent vector exists, it adds
    the coefficient c to the appropriate existing one*) ….]

 Pol["MonomInPol",l_:List,c_:Number]:=Module[{P},   P=Pol["Init"];
    P=Pol["AdaugMonom",P,l,c];  Return[P];   ];

 Pol["Copy",P_]:=Module[(*returns a copy of the polynomial P*) … ]

 Pol["+",P1_,P2_]:=Module[{n1,n2,i,j,k,v1,v2,Rez},
  (* returns the sum of P1, P2 *)
  n1=Pol["Nr",P1]; n2=Pol["Nr",P2]; Rez=Pol["Init"];
  For[i=1;j=1;k=1, k<=n1+n2, k++,
```

```
    v1=Pol["Monom",i,P1]; c1=Pol["Coef",i,P1];
    v2=Pol["Monom",j,P2]; c2=Pol["Coef",j,P2];
  If[!(DVect["=",v1,DVect["Max"]] && DVect["=",v2,DVect["Max"]]),
    If[DVect["=",v1,v2],
       Rez=Pol["AdaugMonom",Rez,v1,DCoef["+",c1,c2]]; i++;j++,
       If[DVect["<",v1,v2],
          Rez=Pol["AdaugMonom",Rez,v1,c1]; i++,
          Rez=Pol["AdaugMonom",Rez,v2,c2]; j++]]; ,] ];
  Print[Pol["Out",Rez]]; Return[Rez]; ];

 Pol["*",P1_:List,P2_:List]:=Module[{n1,n2,v1,v2,i,j,Rez},
  (* returns the product of P1, P2 *)
  n1=Pol["Nr",P1]; n2=Pol["Nr",P2]; Rez=Pol["Init"];
  For[i=1, i<=n1, i++,
     v1=Pol["Monom",i,P1]; c1=Pol["Coef",i,P1];
     For[j=1, j<=n2, j++,
        v2=Pol["Monom",j,P2]; c2=Pol["Coef",j,P2];
        Rez=Pol["AdaugMonom", Rez, DVect["+",v1,v2],
                     DCoef["*",c1,c2]];    ]];
  Print[Pol["Out",Rez]]; Return[Rez]; ];

  Pol["&",nr_:Number,P_:List]:=Module[(*multiplies P by the number n*)…]

  Pol["-",P1_,P2_]:=Module[{P}, P=Pol["&",-1,P2];
Return[Pol["+",P1,P]];   ];

  Pol["/",P_,v_List,c_:Number]:=Module[…  (* divides each of P's
monomials by
     the exponent vector v and coefficient c and returns the result *) …
]
  Pol["|",v_List,P_]:=Module[(* tests whether the exponent vector v
divides any
      of P's monomials and returns True or False *) …]

  Pol["Out",P_]:=Module[{i,nr,l}, (*polynomial display*)
   nr=Pol["Nr",P]; l={};
   For[i=1,i<=nr,i++,
     l=Append[l,Pol["Coef",i,P]*DVect["Out",Pol["Monom",i,P]]]];
   Return[Apply[Plus,l]]];

  Pol["Inp",e__]:=Module[…  (*transforms an input polynomial into the
internal
   form; the code is rather complex and based on Mathematica internal
forms*) …]
] (*Module*)

End[]
EndPackage[]
```

### 2.3. Defining the Coefficient Domains

With the view to revealing the practical use of the categorical definition for creating various polynomial domains, we experimentally define, pursuing the same parameterization principles, the following coefficient domains, derived from the abelian ring category: the integer domain Z (DomZ function), the integer mod n domain, for prime n – `Zmod` and the square matrices domain, with symbolic or numerical elements (`Matrice` function). More rigorously, matrices can be defined over a specific coefficient domain, but this aspect is not of interest within the present application. In order to use `Z` and `Zmod` domains for Gröbner bases algorithms, we define them as integrity and Gcd domains. We observe that the extension of the mod n remainder classes ring with a `Quotient` operator is consistent for prime n but polynomial operations can be performed in `Z` mod n for any natural n. We give below the Mathematica code which defines the above described domains.

```
DomZ[Z_]:=Module[{}, InelCom[Z,"+","*","0","1"];(*abelian ring[And97]*)
  Z["+",a_Integer,b_Integer]:=a+b;
  Z["-",a_Integer,b_Integer]:=a-b;
  Z["*",a_Integer,b_Integer]:=a*b;
  Z["Cat",a_Integer,b_Integer]:=Quotient[a,b]; (*a/b*)
  Z["Mod",a_Integer,b_Integer]:=Mod[a,b];
  Z["GCD",a_Integer,b_Integer]:=GCD[a,b];
  Z["LCM",a_Integer,b_Integer]:=LCM[a,b];
  Z["0"]:=0; (*neutral element*)
  Z["=", a__, b__]:=SameQ[a,b];
  Z["<>", a__, b__]:=UnsameQ[a,b];
  Z["/",a_List,b_Integer]:=a/b;  ]

Zmod[Zm_,n_Integer]:=Module[{n1,n2},
  InelCom[Zm,"+","*","0","1"];                  (*abelian ring [And97]*)
  Zm["+",n1_Integer,n2_Integer]:=Mod[n1+n2,n];
  Zm["-",n1_Integer,n2_Integer]:=Mod[n+n1-n2,n];
  Zm["*",n1_Integer,n2_Integer]:=Mod[n1*n2,n];
  Zm["Catm",n1_Integer,n2_Integer]:=Mod[Quotient[n1+n,n2],n];
  Zm["Cat",n1_Integer,n2_Integer]:=Module[{x},
    For[x=0,x<=n-1,x++,    If[Zm["*",n2,x]==n1, Return[x],]];
    Print["Impartire cu 0"];Return[-1]; (*division by 0*)  ];
  Zm["/",n1_Integer,n2_Integer]:=Zm["Cat",n1,n2]; (*Mod[n1/n2,n];*)
  Zm["/",a_List,b_Integer]:=Module[{i,l},    l={};
    For[i=1,i<=Length[a],i++,
      l=Append[l,Zm["/",a[[i]],b]]];  Return[l];  ];
  Zm["Mod",n1_Integer,n2_Integer]:=Module[{c},  c=Zm["Cat",n1,n2];
    If[c!=-1,  Return[Mod[n1-n2*c,n]],
      Print["Impartire cu 0"];Return[-1]] (*division by 0*)  ];
  Zm["GCD",n1_Integer,n2_Integer]:=Module[{r,a,b},
    r=Zm["Mod",n1,n2];a=n1;b=n2;
    While[r!=0,  a=b; b=r; r=Zm["Mod",a,b];  ];   Return[a];  ];

Zm["LCM",n1_Integer,n2_Integer]:=Zm["Cat",Zm["*",n1,n2],Zm["GCD",n1,n2]
];
  Zm["0"]:=0;
  Zm["=", a__, b__]:=SameQ[a,b];
  Zm["<>", a__, b__]:=UnsameQ[a,b];];

Matrice[Mat_,n_Integer]:=Module[{m1,m2}, InelCom[Mat,"+","*"];
  Mat["+",m1_List,m2_List]:=m1+m2;
  Mat["*",m1_List,m2_List]:=Dot[m1,m2];
  Mat["Inv",m_List]:=Inverse[m];
  Mat["Transp",m_List]:=Transpose[m];
```

```
Mat["Det",m_List]:=Det[m];
Mat["0"]:=Module[{i,l,ll},  (*neutral element*)
 For[l={};i=1,i<=n,i++,      l=Append[l,0]];
 For[ll={};j=1,j<=n,j++,     ll=Append[ll,l]];
 Return[l];  ]; ];
```

We studied the efficiency of our implementation versus the built-in one; the latter is obviously more efficient for common operations but it does not allow polynomial operations over the new implemented domains, i. e. integers mod n and matrices. Moreover, our implementation is *easily extendable* for new coefficient domains by using the same definition of the multivariate polynomial category. Further execution details, referring to run times in Mathematica 3.0 Kernel [Wol92], on a 16MB RAM, 100MHz Pentium, are given in the following table.

It can be noticed that the efficiency of **polinom.m** and **polin.m** implementations is similar; though, for some cases, referring table elements proves to be faster than selecting list elements. Nevertheless, **polin.m**'s representation has certain drawbacks regarding Mathematica parameter transmission: a polynomial is transmitted by the corresponding table symbol, fact that can create inconsistencies when table elements are modified. For this reason, we use **polinom.m** implementation within the Gröbner bases algorithms which are described in the following section.

| Polynomial domain | $Z[x,y,z]$ | | | | | | $Z_7[x,y,z]$ | | | | | | $M_2[x,y,z]$, $M_2 = 2 \times 2$ (numeric or symbolic) matrix domain | | | |
|---|---|---|---|---|---|---|---|---|---|---|---|---|---|---|---|---|
| Operation (P1 "o" P2) | Sum | | | Product | | | Sum | | | Product | | | Sum | | Product | |
| Number of terms | 2 | 7 | 10 | 2 | 7 | 10 | 2 | 7 | 10 | 2 | 7 | 10 | 2 | 5 | 2 | 5 |
| *Mathematica Implementation* | < .01 s | < .01 s | < .01 s | .06 s | < .01 s | .06 s | _ | _ | _ | _ | _ | _ | _ | _ | _ | _ |
| *Implementation 1 –polinom.m package* | .11 s | .27 s | .49 s | .05 s | 2.2 s | 8 s | .11 s | .28 s | .49 s | .11 s | 2.3 s | 6.8 s | .16 s | .44 s | 16 s | 1.5 s |
| *Implementation 2 –polin.m package* | .16 s | .22 s | .66 s | .05 s | 2 s | 7.2 s | .22 s | .28 s | .44 s | .16 s | 2 s | 6.4 s | .11 s | .22 s | .06 s | .22 s |

### 3. Implementing Gröbner Bases Algorithms

We implemented Buchberger's algorithms for computing the Gröbner basis and the reduced Gröbner basis [Buc85] of a polynomial set into Mathematica packages: **groebner.m** and **groebred.m**. The functions which compute the Gröbner bases are parameterized with a polynomial domain, therefore they can be applied for polynomial domains over any consistent coefficient domain that is previously defined. Note that a polynomial domain is created by using the polynomial categorical definition within **polinom.m** package (see 2.2), which is parameterized with a coefficient domain defined in **domcoef.m** package (see 2.3).

Within **groebner.m** package we implemented Buchberger's Gröbner basis algorithm [Buc85] – `BazaGroebner[…]` function. We completely described the algorithmic iterations for computing the normal form of a polynomial modulo a polynomial set – `Normal[Pol,F,g]` function, where `Pol` is the current polynomial domain. For computing the S-polynomial of two polynomials, we implemented the formula proposed in [Gru93] – `SPol[…]` function.

In the code given below, one can noticed the specific functional and parameterized syntax used for expressing operations over various domains (exponent vector, polynomial or coefficient).

```
BeginPackage["Groebner`"]

Normal::usage="Normal[Pol,F,g] verifica daca g este in forma
  normala modulo F, in domeniul de polinoame Pol"
FormaNormala::usage="FormaNormala[Pol,DCoef,DVect,F,p] returneaza forma
  normala a lui p modulo F; operatiile se efectueaza in domeniul de
  polinoame Pol"
SPol::usage="SPol[Pol,DCoef,DVect,P1,P2] calculeaza, in domeniul de p
   olinoame Pol(DCoef,DVect), Rez=SPol(P1,P2)"
BazaGroebner::usage="BazaGroebner[Pol,DCoef,DVect,F] returneaza baza
  Groebner a multimii de polinoame F, in domeniul de polinoame
  Pol(DCoef,DVect)"
MultPolExtInInt::usage="MultPolExtInInt[Pol,M] transforma multimea
  de polinoame din reprezentare externa intr-o multime in care
  polinoamele sunt in format intern (operatii in domeniul Pol)"
MultPolIntInExt::usage="MultPolIntInExt[Pol,M] transforma multimea
  de polinoame din reprezentare interna intr-o multime in care
  polinoamele sunt in format extern (operatii in domeniul Pol)"
Tiparire::usage="tipareste o multime de polinoame reprezentate
  in forma interna"
TipPerechi::usage="tipareste o multime de perechi de polinoame
  reprezentate in forma interna"

Begin["Groebner`Private`"]
Needs["Polinom`"]

PolNormal[Pol_,F_List,g_]:=Module[{norm,i,dim},
 (*Verifies whether g is in normal form mod F, i. e. no monomial of g
  is divisible by the "leading  monomial of any polynomial belonging to
  F – set of polynomials. Operations are performed within the
  polynomial domain Pol*)
  norm=True;  dim=Length[F];
  For[i=1, i<=dim, i++,
    If[Pol["|",Pol["MonomGrMax",F[[i]]],g], norm=False,]];
  Return[norm];]

FormaNormala[Pol_,DCoef_,DVect_,F_List,p_]:=Module[
  {fi,monmaxi,coefmaxi,x,cx,pp,aux,u,alfa,beta,rr,vv,j,dd,afi,rez},
```

```
(*Returns p's normal form mod F; operations are performed within the
 polynomial domain Pol(DCoef,DVect) *)
  rez=Pol["Copy",p];
  While[vv=!PolNormal[Pol,F,rez],
    For[i=1,i<=Length[F],i++,
       fi=Pol["Copy",F[[i]]];    monmaxi=Pol["MonomGrMax",fi];
       coefmaxi=Pol["CoefGrMax",fi];       j=Pol["Nr",rez];
       While[!Pol["Nul",rez] && DVect[">=",
         x=Pol["Monom",j,rez], monmaxi],
        If[DVect["|",monmaxi,x],
         pp=Pol["Copy",rez]; cx=Pol["Coef",j,rez];
         pp[[1]]=Delete[pp[[1]],j];    pp[[2]]=Delete[pp[[2]],j];
         u=DVect["-",x,monmaxi]; (*monomul "cat"*)
         afi=Pol["Copy",fi];   dd=Pol["Nr",afi];
         afi[[1]]=Delete[afi[[1]],dd];   afi[[2]]=Delete[afi[[2]],dd];
         alfa=coefmaxi;    beta=cx;    g=DCoef["GCD",alfa,beta];
         alfa=DCoef["/",alfa,g];beta=DCoef["/",beta,g];
         pp=Pol["&",alfa,pp];    aux=Pol["MonomInPol",u,beta];
         rr=Pol["*",aux,afi];   rez=Pol["-",pp,rr];
         j=Pol["Nr",rez]+1; ];
        j--;      ]    ] ];
   Return[rez];  ]

SPol[Pol_,DCoef_,DVect_,P1_,P2_]:=Module[
   {M1,M2,M,c1,c2,c,P,R1,R2,R,AP1,AP2,Rez},
   (* computes, within the polynomial domain P, Rez=SPol(P1,P2) *)
   M1=Pol["MonomGrMax",P1]; c1=Pol["CoefGrMax",P1];
   M2=Pol["MonomGrMax",P2]; c2=Pol["CoefGrMax",P2];
   M=DVect["LCM",M1,M2];    c=DCoef["LCM",c1,c2];
   P=Pol["MonomInPol",M,1];  AP1=Pol["&",c,P1];  AP2=Pol["&",c,P2];
   R1=Pol["/",AP1,M1,c1];  R2=Pol["/",AP2,M2,c2]; R=Pol["-",R1,R2];
   Rez=Pol["*",P,R];  Return[Rez];  ]

MultPolIntInExt[Pol_,M_List]:=Module[{i,N},
(*transforms the polynomial set M from internal form into a set
  containing external forms*)    N={};
  For[i=1,i<=Length[M],i++,
     N=Append[N,Pol["Out",M[[i]]]];];
   Return[N];]

MultPolExtInInt[Pol_,M_List]:=Module[{i,N,f},
   (*transforms the polynomial set M
   from external form into a set containing internal forms *)
  N={};
  For[i=1,i<=Length[M],i++,  Pol["Inp",M[[i]],f];  N=Append[N,f]; ];
  Return[N];]

Tiparire[Pol_,M_List]:=Module[{i}, (*displays a polynomial set*) …]
TipPerechi[Pol_,M_List]:=Module[(*displays a set of polynomial pairs*)]

BazaGroebner[Pol_,DCoef_,DVect_,baza_List,M_List]:=Module[
   {i,j,nrpol,B,f,fi,f1,f2,G,GE,h,hn,hh,l,P},
   (* returns the Groebner basis of the polynomial set M (given as a
      list), within the polynomial domain Pol *)  nrpol=Length[M];
   Polinom[Pol,DCoef,DVect,baza]; (*creates the polynomial domain -2.2*)
   MultPolExtInInt[Pol,M,G];
      (*M polynomial set is transformed from the
```

```
                  external form into the internal form G*)
   B={};
   For[i=1, i<=nrpol-1, i++,
     For[j=i+1, j<=nrpol, j++,
       B=Append[B, List[G[[i]],G[[j]]]];  ]];
   While[UnsameQ[B,{}],
     l=B[[1]]; f1=l[[1]]; f2=l[[2]];
     B=Complement[B,{l}];
     h=SPol[Pol,DCoef,DVect,f1,f2];
     hn=FormaNormala[Pol,DCoef,DVect,G,h];
     If[!Pol["Nul",hn] (*UnsameQ[Pol["Out",hn],0]*),
       For[i=1, i<=Length[G], i++,  B=Append[B,{G[[i]],hn}]];
       G=Append[G,hn] ,   ]   ];
         (*G polynomial set is transformed from the
         internal form into the external form*)
   MultPolIntInExt[Pol,G,GE];    Tiparire[G];    Return[GE]; ]

End[]
EndPackage[]
```

**Groebred.m** package implements Buchberger's algorithm for computing the reduced Gröbner basis [Buc85]. Applied on a set of three polynomials with 6-7 terms, its run time varies between 40 and 50 seconds, on the same 16MB RAM, 100 MHz Pentium computer.

## 4. Conclusions

The domain / category theory in symbolic computation combines algebraic and programming principles with the view to modeling symbolic computations within various algebraic structures. In this theoretical context, we propose a few definitions for some type theory concepts, by unifying the algebraic and object oriented points of view.

The categorical and parameterized approach reveals a general method for defining and easily extending new computational domains (in this case, not implemented in a classical SCS): the same categorical definition can be parameterized with any specific domain, once defined. In the mean time, this approach provides appropriate protection to inconsistent operations over the defined domains, since it simulates OO principles.

The extension of Mathematica, a widely used symbolic computation system, with a new type system, including algebraic structures and polynomial domains, enables the user to perform operations within new working contexts, such as polynomial domains over various coefficient fields.

## 5. Acknowledgements

I thank Prof. Bazil Pârv from the Computer Science Department, "Babeş-Bolyai" University, for initiating my work regarding practical applications of algebraic category definitions.